\documentclass[12pt, a4paper]{article}
\usepackage{a4}
\usepackage{amsmath,amssymb}
\usepackage{graphicx}
\usepackage{subfigure}
\usepackage{color}
\usepackage[bf,footnotesize]{caption}

\graphicspath{{plots/}}
\def \3{\ss }

\newcommand{\tr}{\operatorname{Tr}}
\newcommand{\re}{\operatorname{Re}}

\newcommand{\beq}{\begin{equation}}
\newcommand{\eeq}{\end{equation}}
\newcommand{\beqn}{\begin{eqnarray}}
\newcommand{\eeqn}{\end{eqnarray}}

\def\ors{a}
\def\rmii{b}
\def\mns{c}
\def\val{d}
\def\nic{e}
\def\rmiii{f}
\def\rmi{g}
\def\liv{h}
\def\des{i}
\def\ect{j}
\def\zur{k}

\begin{document}
\begin{titlepage}
  {\vspace{-0.5cm} \normalsize
  \hfill \parbox{60mm}{DESY/06-236, MS-TP-06-34\\
                       RM3-TH/07-1, ROM2F/2007/02\\
                       SFB/CPP-06-57}}\\[10mm]
  \begin{center}
    \begin{LARGE}
      \textbf{Dynamical Twisted Mass Fermions with Light Quarks} \\
    \end{LARGE}
  \end{center}

  \vskip 0.5cm
  \begin{figure}[h]
    \begin{center}
      \includegraphics[draft=false]{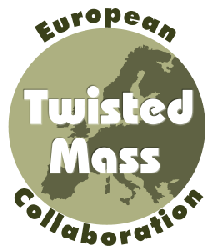}
    \end{center}
  \end{figure}

  \vspace{-0.8cm}
  \baselineskip 20pt plus 2pt minus 2pt
  \begin{center}
    \textbf{
      Ph.~Boucaud$^{(\ors)}$,
      P.~Dimopoulos$^{(\rmii)}$,
      F.~Farchioni$^{(\mns)}$,
      R.~Frezzotti$^{(\rmii)}$,
      V.~Gimenez$^{(\val)}$,
      G.~Herdoiza$^{(\rmii)}$,
      K.~Jansen$^{(\nic)}$,
      V.~Lubicz$^{(\rmiii)}$,
      G.~Martinelli$^{(\rmi)}$,
      C.~McNeile$^{(\liv)}$,
      C.~Michael$^{(\liv)}$,
      I.~Montvay$^{(\des)}$,
      D.~Palao$^{(\val)}$,
      M.~Papinutto$^{(\rmiii)}$,
      J.~Pickavance$^{(\liv)}$,
      G.C.~Rossi$^{(\rmii)}$,
      L.~Scorzato$^{(\ect)}$,
      A.~Shindler$^{(\nic)}$,
      S.~Simula$^{(\rmiii)}$,
      C.~Urbach$^{(\liv)}$,
      U.~Wenger$^{(\zur)}$}\\
  \end{center}
  
  \begin{center}
    \begin{footnotesize}
      \noindent 

      $^{(\ors)}$ Laboratoire de Physique Th\'eorique (B\^at.~210), Universit\'e
      de Paris XI,\\ Centre d'Orsay, 91405 Orsay-Cedex, France\\
      \vspace{0.2cm}

            $^{(\rmii)}$ Dip. di Fisica, Universit{\`a} di Roma Tor Vergata and INFN,
      Sez. di Roma Tor Vergata,\\ Via della Ricerca Scientifica, I-00133 Roma, Italy\\
      \vspace{0.2cm}
      
      $^{(\mns)}$ Universit\"at M\"unster, Institut f\"ur Theoretische Physik,
      \\Wilhelm-Klemm-Strasse 9, D-48149 M\"unster, Germany\\
      \vspace{0.2cm}
      
      $^{(\val)}$ Dep. de F\'{\i}sica Te\`{o}rica and IFIC, Univ. de Val\`{e}ncia,\\
      Dr.Moliner 50, E-46100 Burjassot, Spain\\
      \vspace{0.2cm}
      
      $^{(\nic)}$ NIC, DESY, Zeuthen, Platanenallee 6, D-15738 Zeuthen, Germany\\
      \vspace{0.2cm}
      
      $^{(\rmiii)}$ Dip. di Fisica, Universit{\`a} di Roma Tre and INFN, Sez. di
      Roma III,\\ Via della Vasca Navale 84, I-00146 Roma, Italy\\
      \vspace{0.2cm}
      
      $^{(\rmi)}$ Dip. di Fisica, Universit\`a di Roma ``La Sapienza'',
      and INFN, Sezione di Roma, \\P.le A.~Moro 2, I-00185 Rome, Italy\\
      \vspace{0.2cm}
      
      $^{(\liv)}$ Theoretical Physics Division, Dept. of Mathematical Sciences,
      \\University of Liverpool, Liverpool L69 7ZL, UK\\
      \vspace{0.2cm}
      
      $^{(\des)}$ Deutsches Elektronen-Synchrotron DESY, Notkestr.\,85, D-22607
      Hamburg, Germany\\
      \vspace{0.2cm}
      
      $^{(\ect)}$ ECT* Strada delle Tabarelle 286, I-38050 Villazzano (TN), Italy
      \vspace{0.2cm}
      
      $^{(\zur)}$ Institute for Theoretical Physics, ETH Z{\"u}rich, CH-8093 Z{\"u}rich,
      Switzerland\\
      
    \end{footnotesize}
  \end{center}
  
\end{titlepage}

  \begin{abstract}
    \noindent We present results of dynamical simulations of $N_f=2$ degenerate Wilson  
    twisted mass
    quarks at maximal twist in the range of pseudo scalar masses 
    $300\ \mathrm{MeV} \lesssim m_\mathrm{PS} \lesssim 550\ \mathrm{MeV}$. 
    Reaching such small masses was made possible owing to a recently developed
    variant of the HMC algorithm.  
    The simulations are performed at one value of the lattice spacing 
    $a\lesssim 0.1\ \mathrm{fm}$. 
    In order to have $\mathcal{O}(a)$ improvement and aiming at 
    small residual 
    $\mathcal{O}(a^2)$ cutoff effects, the theory is tuned to maximal twist by 
    requiring the vanishing of the untwisted quark mass.
    Precise results for the pseudo scalar decay constant and the 
    pseudo scalar mass are confronted with chiral perturbation theory 
    predictions and the low energy constants $F$, $\bar{l}_3$ and 
    $\bar{l}_4$  are evaluated with small statistical errors.
  \end{abstract}

\section{Introduction}

The Wilson twisted mass formulation of lattice QCD, though a rather
recent approach, has been by now well established. It amounts to
adding a twisted mass term to the standard, un-improved Wilson-Dirac
operator \cite{Wilson:1974sk} leading to so-called Wilson twisted mass
fermions \cite{Frezzotti:2000nk,Frezzotti:2003ni}. 

Besides being a theoretically sound formulation of lattice QCD, 
Wilson twisted mass fermions offer a number of advantages when tuned to
maximal twist: (i) in this case
automatic $\mathcal{O}(a)$ improvement is obtained by tuning only one
parameter, the bare untwisted quark mass, while avoiding additional tuning of
operator-specific improvement-coefficients; (ii) the mixing
pattern in the renormalisation process  can be 
significantly simplified; (iii)
the twisted mass provides an infra-red regulator helping to overcome 
possible problems with ergodicity in 
molecular dynamics based algorithms\footnote{Although in the light of 
recent algorithmic developments
\cite{Hasenbusch:2001ne,Peardon:2002wb,AliKhan:2003br,Luscher:2004rx,Urbach:2005ji,Clark:2006fx}
this property does not seem to be that important anymore, we consider
it still to be an advantage to have an infra-red regulator  
in the theory which helps in stabilising the simulations. For a recent
stability analysis of pure Wilson fermion simulations see
Ref.~\cite{DelDebbio:2005qa}}.

In the {\em quenched approximation}, these expectations 
--~based on general field theoretical and chiral perturbation 
theory ($\chi$PT) related arguments  
\cite{Frezzotti:2000nk,Frezzotti:2003ni,Frezzotti:2004wz,Frezzotti:2005gi,Sharpe:2004ny,Sharpe:2005rq,Aoki:2004ta}~--
could be verified in actual simulations 
\cite{Jansen:2003ir,Jansen:2005gf,Jansen:2005kk,Abdel-Rehim:2005gz}:     
$\mathcal{O}(a)$ improvement is indeed realised when the theory is 
tuned to maximal twist. 
Moreover, it has been shown that a particular realisation of maximal 
twist, requiring 
parity restoration, also suppresses the $\mathcal{O}(a^2)$ cut-off effects 
substantially, even 
at small quark masses corresponding to values of the pseudo scalar mass of 
$m_\mathrm{PS}\lesssim 300$MeV. 
In addition, with the twisted mass parameter as an infra-red cut-off in place, 
substantially smaller quark masses could be obtained, compared to those 
reachable by standard or 
$\mathcal{O}(a)$ improved Wilson fermions which are plagued by 
so-called exceptional configuration problems in the quenched approximation.
In Refs.~\cite{Pena:2004gb,Guagnelli:2005zc,Dimopoulos:2006dm}  
it was shown that ``wrong chirality'' mixing effects
in the renormalisation process are substantially reduced. In Ref. \cite{Frezzotti:2004wz}
it was proved that all such mixings can be eliminated if a mixed 
action with maximally twisted sea quarks and appropriately chosen 
Osterwalder--Seiler valence fermions is employed.
For a further discussion of the potential of twisted mass QCD on the lattice, 
see Refs.~\cite{Becirevic:2006ii,Chiarappa:2006ae,Ilgenfritz:2006tz,Bar:2006zj}.

The main drawback of the twisted mass approach is the explicit breaking 
of parity and isospin symmetry which are only restored when the 
continuum limit is reached. 
However, due to $\mathcal{O}(a)$ improvement, this 
breaking is an $\mathcal{O}(a^2)$ effect as confirmed
by simulations performed in the quenched approximation 
\cite{Jansen:2005cg,Farchioni:2005hf}.
For recent reviews of the status of Wilson twisted mass fermions see
Refs.~\cite{Farchioni:2005ec,Scorzato:2005rb,Shindler:2005vj} 
and references therein.

It is the main goal of our collaboration to compute a number of
phenomenologically relevant quantities with \emph{dynamical quarks},  
\emph{in the continuum limit} 
and at \emph{small values of the pseudo scalar mass}. 
As a first step in this direction we  here present results for $N_f=2$ 
mass-degenerate quarks at a fixed lattice spacing 
$a\lesssim 0.1\ \mathrm{fm}$. 
We have so far concentrated on the pseudo scalar mass $m_\mathrm{PS}$,
covering a range of values $300\mathrm{MeV} \lesssim m_\mathrm{PS} \lesssim
550\mathrm{MeV}$, the pseudo scalar decay
constant $f_\mathrm{PS}$ and the static inter-quark force parameter $r_0$ at five values
of the quark mass. A wider range of physical observables will be
addressed in the future.
The results for $m_\mathrm{PS}$ and $f_\mathrm{PS}$ are  
confronted with predictions of $\chi$PT which 
allows extracting the low-energy constants $\bar{l_3}$,
$\bar{l_4}$, $F$ and $B_0$ of the corresponding effective chiral 
Lagrangian. 
We also provide a determination of the size of 
isospin violation measured from the mass splitting
between the lightest charged and neutral pseudo scalar particles. 
First accounts of our work were presented at recent 
conferences, see Refs.~\cite{Jansen:2006rf,Shindler:2006tm}.
In this publication we will focus on the results of our 
present simulations obtained at one value of $\beta$ and 
one volume. We shall provide, in a forthcoming paper \cite{etmcfollow}, 
a comprehensive description of our analysis procedure 
and address systematic errors by including future 
runs on larger lattices, at different values of $\beta$ 
and with extended statistics. Related works with Wilson fermions at
similar small pseudo scalar meson masses are published in
Refs.~\cite{DelDebbio:2006cn,DelDebbio:2007pz,Gockeler:2006ns}.

\section{Choice of Lattice Action}

The Wilson twisted mass 
fermionic lattice action for two flavours of degenerate
quarks reads (in the so called twisted basis
\cite{Frezzotti:2000nk} and fermion fields with continuum dimensions)
\begin{equation}
  \label{eq:Sf}
  \begin{split}
    S_\mathrm{tm} = &\, a^4\sum_x\Bigl\{
    \bar\chi_x\left[m_0 + i\gamma_5\tau_3\mu + \frac{4r}{a}\right]\chi_x\Bigr. \\
    & \Bigl. +\frac{1}{2a}\sum_{\nu=1}^4
    \bar\chi_x\left[U_{x,\nu}(-r+\gamma_\nu)\chi_{x+\hat\nu} +
      U_{x-\hat\nu,\nu}^\dagger(-r-\gamma_\nu) \chi_{x-\hat\nu}\right]\Bigr\}\, ,
  \end{split}
\end{equation}
where $am_0$ is the bare untwisted quark mass and $a\mu$ the bare twisted
mass, $\tau_3$ is the third Pauli matrix acting in flavour space 
and $r$ is the Wilson parameter, which we set to one in our
simulations.
Twisted mass fermions are said to be at {\em maximal twist} if the bare
untwisted mass is tuned to its critical value, $m_\mathrm{crit}$. We
will discuss later how this can be achieved in practice.

In the gauge sector we use the so called tree-level Symanzik improved
gauge action (tlSym) \cite{Weisz:1982zw}, which includes besides the
plaquette term $U^{1\times1}_{x,\mu,\nu}$ also rectangular $(1\times2)$ Wilson loops
$U^{1\times2}_{x,\mu,\nu}$
\begin{equation}
  \label{eq:Sg}
    S_g =  \frac{\beta}{3}\sum_x\Biggl(  b_0\sum_{\substack{
      \mu,\nu=1\\1\leq\mu<\nu}}^4\{1-\re\tr(U^{1\times1}_{x,\mu,\nu})\}\Bigr. 
     \Bigl.+
    b_1\sum_{\substack{\mu,\nu=1\\\mu\neq\nu}}^4\{1
    -\re\tr(U^{1\times2}_{x,\mu,\nu})\}\Biggr)\,  
\end{equation}
with $\beta$ the bare inverse coupling, $b_1=-1/12$ and the
(proper) normalisation condition $b_0=1-8b_1$. Note that at $b_1=0$ this
action becomes the usual Wilson plaquette gauge action.

\subsection{$\mathcal{O}(a)$ improvement}
\label{IMPR}

As mentioned before, $\mathcal{O}(a)$ improvement can be obtained by
tuning Wilson twisted mass fermions to maximal twist. In fact, it was
first proved in Ref.~\cite{Frezzotti:2003ni} that parity even
correlators are free from $\mathcal{O}(a)$ lattice artifacts at maximal
twist by using spurionic symmetries of the lattice action. Later on it
was realised \cite{Frezzotti:2005gi,Shindler:2005vj} that a simpler
proof is possible based on the parity symmetry of the
continuum QCD action and the use of the Symanzik effective theory.

{}From this latter way of proving $\mathcal{O}(a)$ improvement, it
becomes also clear how to define maximal twist: first, choose an operator odd
under parity (in the physical basis) which 
has a zero expectation value in the continuum. 
Second, at a non-vanishing value of the lattice spacing 
tune the expectation value of this operator to zero by
adjusting the value of $am_0$.
This procedure, which has been proposed in 
\cite{Farchioni:2004ma,Farchioni:2004fs} and
has been theoretically analysed in \cite{Frezzotti:2005gi}, 
is sufficient to define maximal twist independently of
the chosen operator. To approach smoothly the 
continuum limit this tuning has to be
performed at fixed physical situation while decreasing the lattice spacing.

It was shown in an extended scaling test in the quenched approximation,
that $\mathcal{O}(a)$ improvement works extremely well for maximally
twisted mass quarks
\cite{Jansen:2003ir,Jansen:2005gf,Jansen:2005kk}. In the context of
this scaling test, the method of setting the so-called PCAC mass to zero 
was found to be very successful, in agreement with
theoretical considerations \cite{Aoki:2004ta,Sharpe:2004ny,Frezzotti:2005gi}. 
 Here the PCAC mass
\begin{equation}
  \label{eq:mpcac}
  m_\mathrm{PCAC} = \frac{\sum_\mathbf{x}\langle\partial_0 A_0^a(x)
    P^a(0)\rangle} {2\sum_\mathbf{x}\langle P^a(x) P^a(0)\rangle}  \, ,
 \qquad a=1,2 \, 
\end{equation}
 is evaluated at large enough time separation, such that the pion 
ground state is dominant.  
To see that the procedure of defining $am_\mathrm{crit}$ from the
vanishing of $m_\mathrm{PCAC}$ is the appropriate one, it is enough to
recall that under that condition the multilocal operator
$\sum_\mathbf{x}A_0^a(x)P^a(0)$ becomes, in the physical basis, 
 the parity odd operator 
$\epsilon^{3ab}\sum_\mathbf{x}\bar\psi\gamma_0\tau^b\psi(x)\,\bar\psi\gamma_5\tau^a\psi(0)$.

In principle one could think of determining $am_\mathrm{crit}$ at each value 
of $a\mu$ at which simulations are performed, possibly followed 
by an extrapolation to vanishing $a\mu$.
 The strategy we are following in this paper is, instead,  to take the
value of $am_\mathrm{crit}$ from the simulation  at the lowest available
value $a\mu_\mathrm{min}\ll a\Lambda_\mathrm{QCD}$.  In this situation $\mathcal{O}(a)$
improvement is still guaranteed, because working at $\mu_\mathrm{min}$ 
merely leads to $\mathcal{O}(a\mu_\mathrm{min}\Lambda_\mathrm{QCD})$ effects in
$m_\mathrm{crit}$ and
$\mathcal{O}(a^2\mu_\mathrm{min}\Lambda_\mathrm{QCD})$ relative corrections in
physical quantities~\cite{Frezzotti:2005gi}. 

\subsection{Phase Structure}
\label{PHST}

In order to understand our choice of the gauge action,
it is important to realise that  
Wilson-type fermions have a non-trivial phase structure at finite
lattice spacing: 
in a series of publications 
\cite{Farchioni:2004us,Farchioni:2004ma,Farchioni:2004fs,Farchioni:2005ec,Farchioni:2005bh,Farchioni:2005tu}
the phase structure of lattice QCD was explored. For
lattice spacings $a \geq 0.15\ \mathrm{fm}$ clear signals
of first order phase transitions at the chiral point were found. 
The strength of those phase transitions 
weakens when the continuum limit is approached. 
This phase transition was identified to be a \emph{generic}
property of Wilson-type fermions since the phenomenon takes place for
the pure Wilson as well as the Wilson twisted mass formulation 
\cite{Frezzotti:2000nk,Frezzotti:2003ni} of lattice QCD. 
Also the properties of physical quantities measured in both 
metastable branches of this first order phase transition were studied
and compared to results of $\chi$PT
\cite{Sharpe:2004ny,Aoki:2004ta,Sharpe:2004ps,Munster:2004am,Munster:2004wt,Scorzato:2004da} 
finding that (lattice) $\chi$PT describes the simulation 
data quite well.
 This is somewhat surprising since the simulation data were obtained  at
rather coarse values of the lattice spacing and at rather heavy 
pseudo scalar 
masses, where the applicability of $\chi$PT  may
be questionable.

A very important consequence of the first order phase transition 
phenomenon is that at non-vanishing lattice spacing, simulations
cannot be performed with pseudo scalar mesons below a
certain minimal mass value. 
From lattice $\chi$PT analyses it is expected that this minimal value 
of the pseudo scalar mass goes to zero with a rate of
$\mathcal{O}(a)$. 
In different words, given a value of the pseudo scalar mass, $m_\mathrm{PS}$, 
one can always find a value of the lattice spacing 
$a_\mathrm{max}(m_\mathrm{PS})$, such that simulations at 
$a<a_\mathrm{max}(m_\mathrm{PS})$ 
can be safely performed.
For example, when the Wilson plaquette gauge action is used one finds
$a_\mathrm{max}\approx 0.07\ \mathrm{fm}$ to realise a pseudo scalar mass of
about $300$MeV \cite{Farchioni:2005tu}.

The phase structure of lattice QCD with Wilson-type fermions has
previously been addressed:
there have been investigations concerning the Aoki-phase 
\cite{Aoki:1986ua} in Refs.~\cite{Sternbeck:2003gy,Ilgenfritz:2003gw} at 
large gauge couplings corresponding to values of the
lattice spacing $a\geq 0.2\ \mathrm{fm}$. 
In other studies \cite{Blum:1994eh,Aoki:2001xq,Aoki:2004iq} signals of 
first order phase transitions were found for Wilson fermions with and without
the clover term, 
see also Ref.~\cite{Jansen:2003nt}.
In Ref.~\cite{Creutz:1996bg} a   
speculative picture of the phase structure of Wilson lattice
QCD has been given and 
in Ref.~\cite{Sharpe:1998xm} an analysis within the framework
of $\chi$PT has been reported.
A detailed understanding of the generic phase structure was obtained 
in the 2-dimensional   
Gross-Neveu model, see Refs.~\cite{Aoki:1985jj,Izubuchi:1998hy,Nagai:2005mi}.
Of course, it is unclear how much these last results are applicable to 
4-dimensional lattice QCD.

In order to choose a gauge action for our production simulations we
studied the phase structure  employing a number of different 
gauge actions: the standard Wilson plaquette gauge action
\cite{Wilson:1974sk} ($b_1 =0$ in Eq.~(\ref{eq:Sg})),
the DBW2 gauge action \cite{deForcrand:1996bx} 
($b_1 = -1.4088$) and the tree-level Symanzik improved
gauge action  \cite{Weisz:1982zw} ($b_1 =-1/12$). A 
marked dependence of the strength of the phase transition 
on the choice of the
gauge action has been found. In particular, these investigations revealed
that the DBW2 and the tlSym gauge actions substantially 
weaken the effect of the
first order phase transition and in particular the value of $a_\mathrm{max}$ 
increases when the coefficient $b_1$ in Eq.~(\ref{eq:Sg}) is moved away 
from zero \cite{Farchioni:2004fs,Farchioni:2005tu}.
We refer to Refs.~\cite{Farchioni:2005ec,Shindler:2005vj}
for summaries of these results.

The DBW2 gauge action 
appears to lead to a bad scaling behaviour
\cite{DeGrand:2002vu,Necco:2003vh,Takeda:2004xh} and a slow
convergence of perturbation theory \cite{Horsley:2004mx}, 
whereas the tlSym gauge action is expected to show
-- by construction -- a good scaling behaviour and a fast convergence of
perturbation theory. Therefore, the tlSym gauge action looks like a
good compromise between the Wilson gauge action
which is most strongly affected by the first order phase transition and
the DBW2 gauge action. 

\section{Numerical Results}

\subsection{Set-up}

\begin{table}[t]
  \centering
  \begin{tabular*}{1.\linewidth}{@{\extracolsep{\fill}}lcccc}
    \hline\hline
    $\Bigl.\Bigr.\beta$ & $L^3\cdot T$ & $a\mu_\mathrm{min}$
    & $\kappa_\mathrm{crit}(a\mu_\mathrm{min})$& $r_0/a$ \\
    \hline\hline
    $3.9$  & $24^3\cdot48$ & $0.004$ & $0.160856$ & $ 5.22(2)$ \\
    \hline\hline
  \end{tabular*}
  \caption{Simulation parameters. We denote by 
    $a\mu_\mathrm{min}$ 
    the smallest value of the twisted mass parameter 
    $a\mu$ at which we have performed
    simulations. At this value of $a\mu$ we determined the critical mass 
    $m_\mathrm{crit}$, or, equivalently
    the critical 
    hopping parameter $\kappa_\mathrm{crit}=1/(8+2am_\mathrm{crit})$. 
    The value of $r_0/a$ has been extrapolated to the physical point, 
    where $m_\mathrm{PS}=139.6$~MeV.}
  \label{tab:setup}
\end{table}

In this letter we will present results at a fixed value 
of the lattice spacing of $a\lesssim 0.1\ \mathrm{fm}$ only. 
In table~\ref{tab:setup} we  
provide the value of $a\mu_\mathrm{min}$ at which we imposed the 
vanishing of $m_\mathrm{PCAC}$ (Eq.~(\ref{eq:mpcac}))
and thus determined
$am_\mathrm{crit}$.  
In table~\ref{tab:results} we list the values of 
the quark mass $am_\mathrm{PCAC}$,
the pseudo scalar mass $am_\mathrm{PS}$, 
the pseudo scalar decay constant $af_\mathrm{PS}$, $r_0/a$ and 
the plaquette integrated autocorrelation time
at all values 
of the twisted mass parameter $a\mu$. All other parameters were kept
fixed as specified in table~\ref{tab:setup}. 

The algorithm we used is a HMC algorithm with mass preconditioning
\cite{Hasenbusch:2001ne,Hasenbusch:2002ai} 
and multiple time scale integration described in detail in
Ref.~\cite{Urbach:2005ji}. The trajectory 
length $\tau$ was set to $\tau=1/2$ in all our runs. Our 
estimates of the plaquette integrated autocorrelation time
$\tau_\mathrm{int}(P)$ quoted in table \ref{tab:results} are in units of
$\tau=1/2$. 
Note that our estimates of the autocorrelation times
of quantities such as $am_\mathrm{PS}$ or $af_\mathrm{PS}$ 
are found to be substantially smaller, typically by a factor of 5-10, 
than those
reported in the table for the plaquette.

As discussed above,  maximal twist is realised in our simulations by
tuning $m_0$ to obtain a vanishing PCAC quark mass $am_\mathrm{PCAC}$ at
the smallest value $a\mu_\mathrm{min}$ of the twisted mass parameter
$a\mu$. From table~\ref{tab:results} one can see that this condition has been
numerically realised with good accuracy, which in this context means
$m_\mathrm{PCAC}(\mu_\mathrm{min})/\mu_\mathrm{min} < a\Lambda_{QCD}$
within statistical errors ($a\Lambda_{QCD} \sim 0.1$ in our case). Once
this is achieved, the (weak) $\mu$-dependence of $m_\mathrm{PCAC}$,  which
is visible in fig. 1(a), is an $\mathcal{O}(a)$ cutoff effect that merely
modifies the $\mathcal{O}(a^2)$ artifacts in physical observables,  as
already mentioned in section 2.1.

In order to make maximum use of the gauge configurations, we evaluate
connected meson correlators using a stochastic method to include all
spatial sources. The method involves a stochastic source (Z(2)-noise in both
real and imaginary part) for  all colour and spatial indices at one
Euclidean time slice. By solving for the propagator from this source  for 
each of the 4
spin components, we can construct zero-momentum meson correlators from
any bilinear at the source and sink. Four inversions of the Dirac matrix 
per Euclidean time slice value are necessary, since we chose to use only one 
stochastic sample per gauge configuration.
This ``one-end''  method is similar to that  pioneered in
Ref.~\cite{Foster:1998vw} and implemented in Ref.~\cite{McNeile:2006bz}.
We also employ a fuzzed source \cite{Lacock:1994qx} of the extent 
of 6 lattice
spacings to enable studying non-local meson creation and destruction
operators.
This allowed us to obtain very stable effective masses and 
to confirm the extraction of the 
pion ground state.

\begin{table}[t]
  \centering
  \begin{tabular*}{1.\linewidth}{@{\extracolsep{\fill}}lccccc}
    \hline\hline
    $\Bigl.\Bigr.a\mu$
    & $am_\mathrm{PS}$ & $af_\mathrm{PS}$ & $am_\mathrm{PCAC}$ &
    $r_0/a$ & $\tau_\mathrm{int}(P)$ \\
    \hline\hline
    $0.0040$ & $0.13587(68)$ & $0.06531(40)$ 
    & $-0.00001(27)$ & $5.196(28)$ & $55(17)$\\
    
    $0.0064$ & $0.16937(36)$ & $0.07051(35)$ & $-0.00009(17)$ & $5.216(27)$ & $23(07)$\\
    
    $0.0085$ & $0.19403(50)$ & $0.07420(24)$ & $-0.00052(17)$ & $5.130(28)$ & $13(03)$\\
    
    $0.0100$ & $0.21004(52)$ & $0.07591(40)$ & $-0.00097(26)$ & $5.143(25)$ &
    $15(04)$\\
    
    $0.0150$ & $0.25864(70)$ & $0.08307(34)$ & $-0.00145(42)$ & $5.038(24)$ &
    $06(02)$\\
    \hline\hline
  \end{tabular*}
  \caption{Results from simulations at $\beta=3.9$ using the simulation
    parameters listed in table~\protect\ref{tab:setup}. The measurements
    were started after $1500$ equilibration trajectories and are based
    on $5000$ equilibrated trajectories.} 
  \label{tab:results}
\end{table}

In general, we save a gauge configuration every second trajectory
and analyse meson correlators as described above from a selection of
different Euclidean time slice sources. To reduce autocorrelations, we
only use the same time slice source every 8-10 trajectories. Our
primary statistical error was obtained with the so called
$\Gamma$-method as described in Ref.~\cite{Wolff:2003sm} and
cross-checked with 
a bootstrap analysis and a jack-knife analysis of blocked data. For a
detailed description of our error analysis we refer to a forthcoming
paper of our collaboration \cite{etmcfollow}.

\begin{figure}[t]
  \centering
  \subfigure[\label{fig:mpcac}]
  {\includegraphics[width=0.45\linewidth]{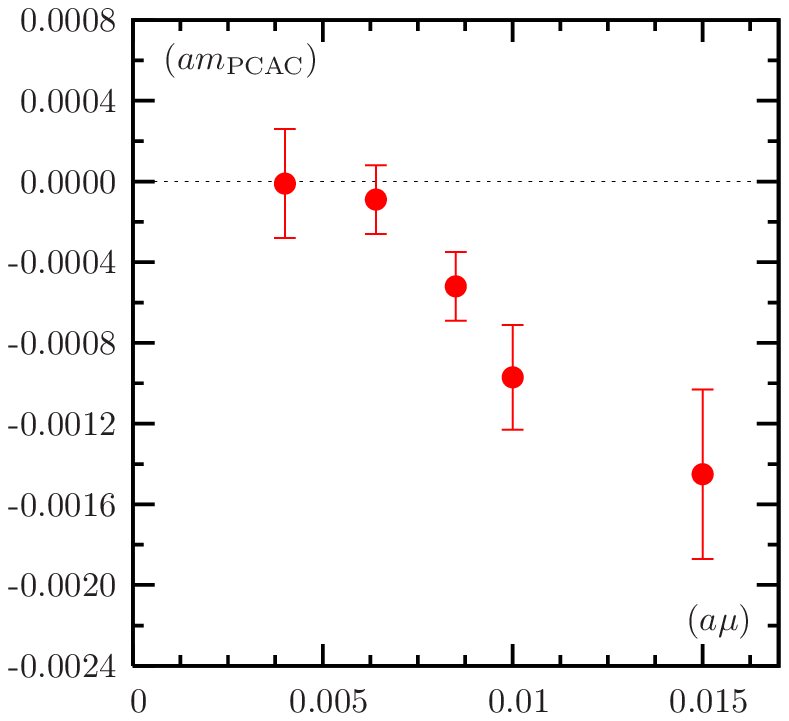}}\quad
  \subfigure[\label{fig:r0}]
  {\includegraphics[width=0.413\linewidth]{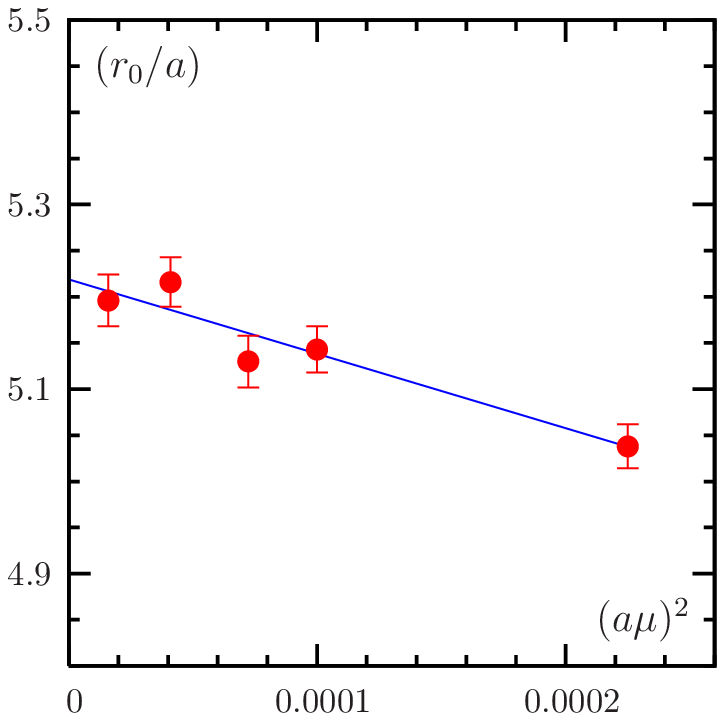}}
  \caption{(a) PCAC quark mass $am_\mathrm{PCAC}$  as function 
    of $a\mu$ and (b) Sommer parameter
    $(r_0/a)$ as functions of $(a\mu)^2$. The solid line 
    in subfigure (b) represents 
    a linear fit in $(a\mu)^2$ to the data.} 
  \label{fig:r0andmpcac}
\end{figure}

\subsection{Force parameter $r_0$}

In simulations of the quenched approximation of lattice QCD, 
the Sommer parameter $r_0$ \cite{Sommer:1993ce}
with a value of $0.5\ \mathrm{fm}$, was widely used to set the 
lattice scale. While $(r_0/a)$ is measurable to good accuracy in lattice QCD
simulations it has the drawback that its value in physical units is
not known very well. Therefore, it becomes necessary to determine the scale
using other quantities which are experimentally accessible with high precision, 
such as $m_\pi$, $f_\pi$, $m_\mathrm{K}$, $f_\mathrm{K}$ or $m_\mathrm{K^*}$.
In fact, in this paper we attempt to determine the lattice scale by
fitting $\chi$PT based formulae 
to our precise data for  $f_\mathrm{PS}$ and $m_\mathrm{PS}$,
using the physical values for  $m_\pi$ and $f_\pi$ as inputs. From this
analysis, we obtain a value of the lattice spacing which is 
10\% lower than  the value  obtained by setting $r_0=0.5\
\mathrm{fm}$. 

Our results for $(r_0/a)$ are reported in table~\ref{tab:results}
and plotted in figure \ref{fig:r0}. Within
the current errors the mass dependence of this quantity appears to be
weak. Since $r_0$ is a pure gauge quantity, it should be 
a function of $(a\mu)^2$ and indeed, a linear fit in $(a\mu)^2$ describes 
the data rather well as shown in figure~\ref{fig:r0}. From the fit we
obtain a value for $r_0/a=5.22(2)$ at the physical point, where 
$a\mu = a\mu_\pi$ (see below), as also quoted in table~\ref{tab:setup}.

\subsection{$f_\mathrm{PS}$ and $m_\mathrm{PS}$ as a function of the
  quark mass}

The {\it charged} pseudo scalar meson mass $am_\mathrm{PS}$ is as usual 
extracted from the time exponential decay of appropriate 
correlation functions~\footnote{Concerning results on the neutral pion mass, 
$am_\mathrm{PS}^0$, see section~(\ref{EIB}).}.
In contrast to pure Wilson fermions, for
maximally twisted mass fermions an exact lattice Ward identity allows
to extract the (charged) pseudo scalar meson decay constant $f_{\rm PS}$ from the
relation 
\begin{equation}
  \label{eq:fps}
  f_\mathrm{PS} = \frac{2\mu}{m_\mathrm{PS}^2} |\langle 0 | P^1 (0)
    | \pi\rangle |\, ,
\end{equation}
with no need to compute any renormalisation constant since 
$Z_P = 1/Z_\mu$ \cite{Frezzotti:2000nk}.
We give our results for $m_\mathrm{PS}$ and $f_\mathrm{PS}$ in 
table~\ref{tab:results}.

We now discuss whether  the continuum $\chi$PT formulae can  reproduce 
the data in table~\protect\ref{tab:results} for $am_\mathrm{PS}$ and
$af_\mathrm{PS}$.
 In our $\chi$PT based analysis, we take into account finite size
corrections because on our lattices at the lowest and next-to-lowest
$\mu$-values they turn out to affect $am_\mathrm{PS}$ and, in particular,  
$af_\mathrm{PS}$ in a significant way.
We have used continuum $\chi$PT to 
describe consistently  the dependence of the data both on the 
finite spatial size ($L$) and on $\mu$.

We fit the appropriate ($N_f=2$) $\chi$PT
formulae~\cite{Gasser:1986vb,Colangelo:2005gd}
\begin{equation}
  m_\mathrm{PS}^2(L) = 2B_0\mu \, \left[
    1 + \frac{1}{2}\xi \tilde{g}_1( \lambda )    \right]^{2} \, \left[ 1 +
    \xi \log ( 2B_0\mu/\Lambda_3^2 ) \right] \, ,
  \label{eq:chirfo1}
\end{equation}
\begin{equation}
  f_\mathrm{PS}(L) = F \, \left[
    1 - 2 \xi \tilde{g}_1( \lambda )    \right] \, \left[ 1 -
    2 \xi \log ( 2B_0\mu/\Lambda_4^2 ) \right] \, ,
  \label{eq:chirfo2}
\end{equation}
to our raw data for $m_\mathrm{PS}$ and $f_\mathrm{PS}$
simultaneously. Here
\begin{equation}
  \xi = 2B_0\mu/(4\pi F)^2 \, , \qquad
  \lambda = \sqrt{2B_0\mu L^2}\ .  
\end{equation}
The finite size correction function $\tilde{g}_1(\lambda)$ was first
computed by Gasser and Leutwyler in Ref.~\cite{Gasser:1986vb} and is
also discussed in Ref.~\cite{Colangelo:2005gd} from which we take our
notation (except that our normalisation of $f_{\pi}$ is 130.7 MeV). 
In Eqs.~(\ref{eq:chirfo1}) and~(\ref{eq:chirfo2}) 
NNLO $\chi$PT corrections are assumed to be negligible.
The formulae above depend on four unknown parameters, 
$B_0$, $F$, $\Lambda_3$ and $\Lambda_4$, which will be determined by the fit. 

We determine  $a\mu_\pi$, the value of $a\mu$ at which the pion 
assumes its physical mass,  by  
requiring that the ratio $[\sqrt{ [m_\mathrm{PS}^2(L=\infty)]
}/f_\mathrm{PS}(L=\infty)]$ takes the value $(139.6/130.7) = 1.068$.
From the knowledge of $a\mu_\pi$ we can evaluate $\bar{l}_{3,4} \equiv
\log(\Lambda_{3,4}^2/m_{\pi}^2)$ and using $f_\pi$ the value of the
lattice spacing $a$ in $\mathrm{fm}$.

In order to estimate the statistical errors affecting our fit values
we generate at each of the
$\mu$-values 1000 bootstrap samples for $m_\mathrm{PS}$ and
$f_\mathrm{PS}$ extracted from the bare correlators, blocked with block-size 
of  32. For each sample (combining all masses) we then fit $m_\mathrm{PS}^2$
and $f_\mathrm{PS}$ simultaneously as a function of $\mu$.
From the 1000 fits we obtain 1000 
bootstrap samples for $2aB_0$, $aF$, $\log(a^2\Lambda_{3,4}^2)$,
$a\mu_\pi$, $a$ and
$\bar{l}_{3,4}$, respectively, from which we compute the corresponding
error estimates, taking in this way the statistical correlation between 
$f_\mathrm{PS}$ and
$m_\mathrm{PS}$ into account.

\begin{figure}[t]
  \centering
  \subfigure[ \label{fig:f1}]
  {\includegraphics[width=0.44\linewidth]{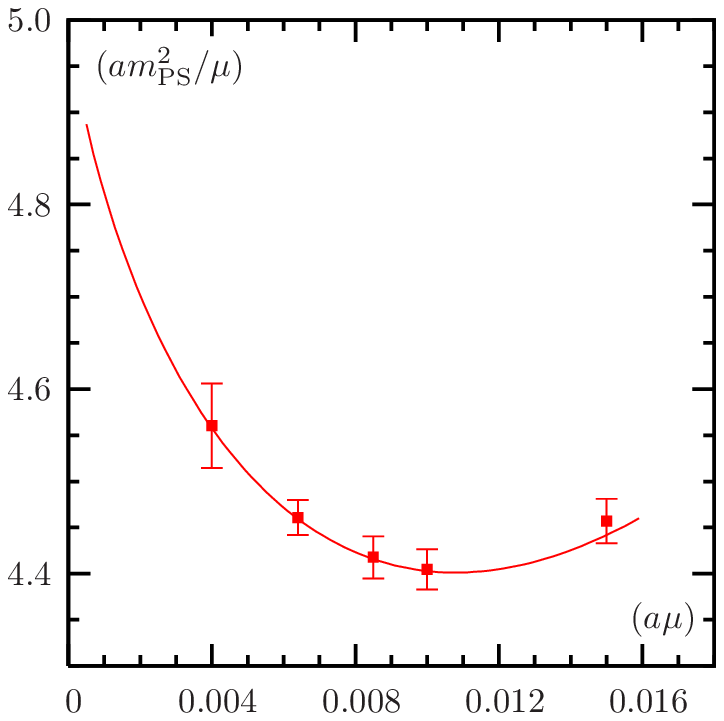}}\quad
  \subfigure[ \label{fig:f2}]
  {\includegraphics[width=0.45\linewidth]{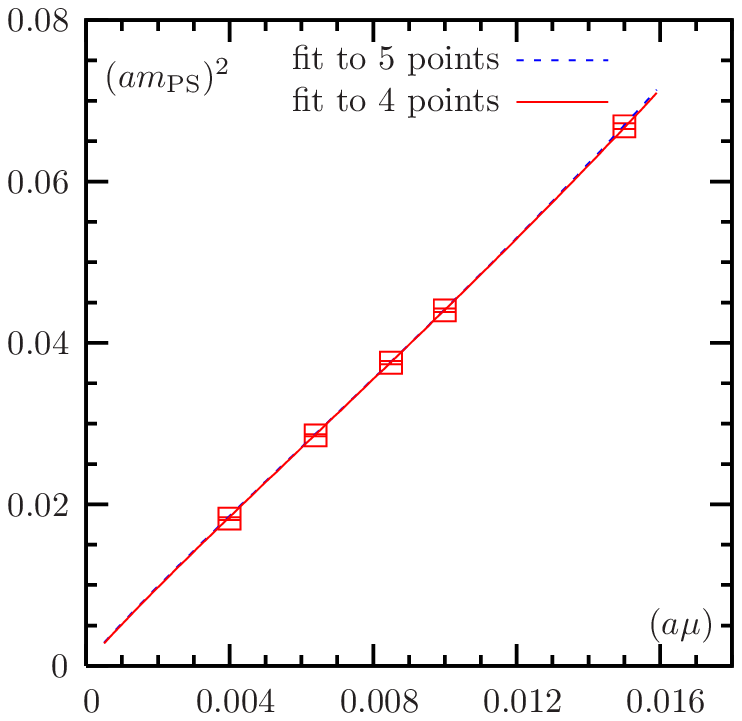}}
  \caption{In (a) we show $(am_\mathrm{PS})^2/(a\mu)$ as a function of
    $a\mu$. In addition we plot the $\chi$PT fit with
    Eq.~(\ref{eq:chirfo1}) to the data from the 
    lowest four $\mu$-values. In (b) we show $(am_\mathrm{PS})^2$ as a
    function of $a\mu$. Here we present two $\chi$PT fits with
    Eq.~(\ref{eq:chirfo1}), one taking all data points and one leaving
    out the point at the largest value  
    $a\mu=0.015$. In both figures (a) and (b) we show finite size
    corrected ($L\to\infty$) data points.}
  \label{fig:mps}
\end{figure}

\begin{figure}
  \centering
  \includegraphics[width=.7\linewidth]{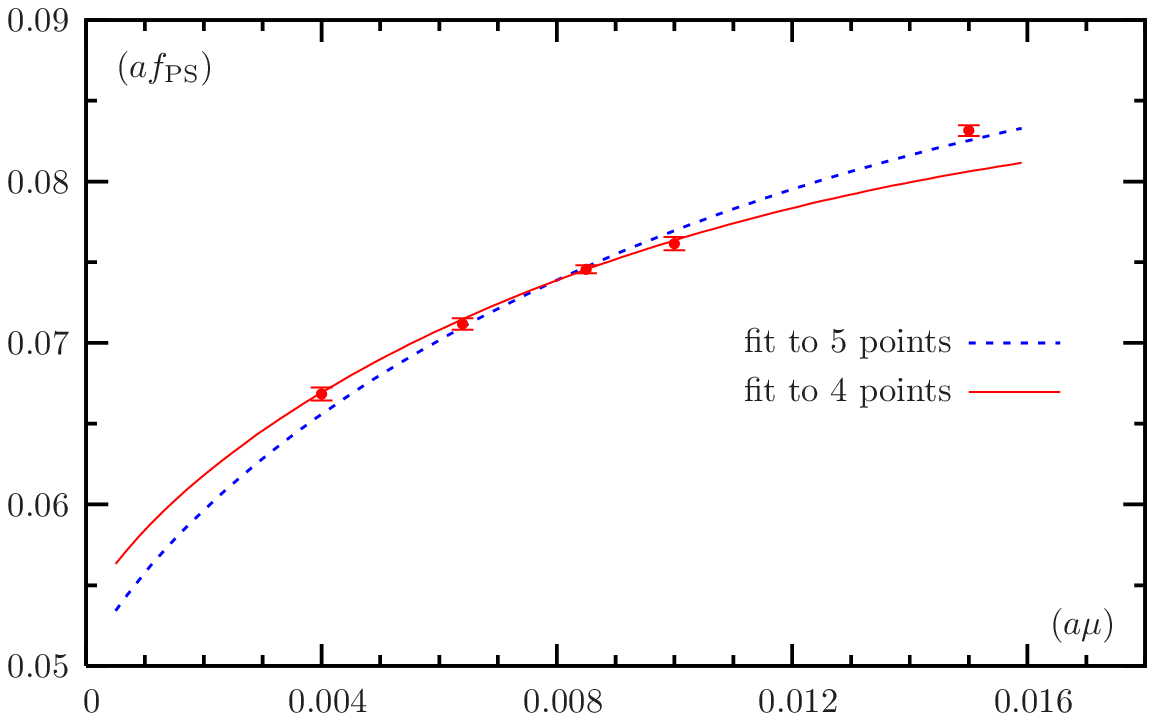}
  \caption{We show $af_\mathrm{PS}$ as a function of $a\mu$ together
    with fits to $\chi$PT formula Eq.~(\ref{eq:chirfo2}). We present
    two fits, one taking all data and one leaving out the point at the
    largest value $a\mu=0.015$. We show finite size corrected
    ($L\to\infty$) data points.}
  \label{fig:f3}
\end{figure}

{}For our lightest four $\mu$-values, we find an excellent fit to our
data on $f_\mathrm{PS}$ and $m_\mathrm{PS}$ (see figures \ref{fig:mps}
and \ref{fig:f3}). The fitted values of the four parameters are
\begin{equation}
  \begin{split}
    2aB_0\quad & =\quad {\color{white}-}4.99(6)\, ,  \\
    aF\quad & =\quad {\color{white}-}0.0534(6)\, ,   \\
    \log(a^2\Lambda_3^2)\quad & =\quad -1.93(10)\, , \\
    \log(a^2\Lambda_4^2)\quad & =\quad -1.06(4)\ . 
    \label{eq:bestfit}
  \end{split}
\end{equation}
Our data are clearly sensitive to $\Lambda_3$ as 
visualised in figure~\ref{fig:f1}. We obtain 
\begin{equation}
  a\mu_\pi = 0.00078(2),\qquad\bar{l}_{3} = 3.65(12), \qquad
  \bar{l}_{4} = 4.52(06)\  
  \label{eq:lec34}
\end{equation}
which compares nicely with other determinations (for a review see
Ref.~\cite{Leutwyler:2006qq}). 
Including also our results from $a\mu=0.0150$ in the fit gives an 
acceptable description of $m_\mathrm{PS}^2$ but misses the data for 
$f_\mathrm{PS}$, as shown in figures~\ref{fig:f2} and~\ref{fig:f3}.
Note, however, that in Eqs.~(\ref{eq:chirfo1}, \ref{eq:chirfo2}), and
thus in the fit results~(\ref{eq:bestfit}, \ref{eq:lec34}), a number of
systematic errors as discussed below are not included.

The values presented here should hence be taken as a first estimate, 
the validity of which will be checked in the future. 
 Nevertheless, the statistical accuracy we are able to achieve  implies
that  there is a very good prospect  of  obtaining   accurate and
reliable values for the low-energy constants from Wilson twisted mass 
fermion simulations.

Since we have obtained an excellent description of our pseudo scalar mesons, 
we can use our fit to extract the lattice spacing. Based on the physical 
value of $f_{\pi}$, we get 
\begin{equation}
  a = 0.087(1)\ \mathrm{fm}\, .
\end{equation}
Using the value of $r_0/a$  reported in table~\protect\ref{tab:setup},
this lattice calibration method yields  $r_0 = 0.454(7)\ \mathrm{fm}$.

We now discuss the possible sources of systematic error.
Our analysis is based on lattice determinations of properties of 
pseudo scalar mesons with masses in the range 300 to 500 MeV 
on lattices with a spatial size slightly above 2~fm. 
Systematic errors can arise from several sources:

\noindent(i) Finite lattice spacing effects.
Preliminary results at a smaller value of the lattice spacing that were 
presented in Refs.~\cite{Jansen:2006rf,Shindler:2006tm}  
suggest that $\mathcal{O}(a)$ improvement is nicely at work and that 
residual $\mathcal{O}(a^2)$ effects are small. 

\noindent (ii) Finite size effects. In order to check that next to leading
order (continuum) $\chi$PT adequately describes these, we are presently 
performing a run at $\beta=3.9$ and $a\mu=0.004$ on a $32^3\cdot 64$ lattice.

\noindent (iii) Mass difference of charged and neutral pseudo scalar 
meson.  
In the appropriate 
lattice $\chi$PT  power-counting for our values of the lattice spacing and quark masses, 
i.e.\ $a\sim \mu\sim p^2$, one gets the order 
of magnitude relation $(m_\mathrm{PS})^2-(m_\mathrm{PS}^0)^2 = 
\mathcal{O}(a^2\Lambda^4_{\mathrm{QCD}}) = \mathcal{O}(p^4)$, from
which it follows that to the order we have been working the effects of
the pion mass splitting do not affect, in particular, 
the finite size correction factors
for $m_{\mathrm PS}$ and $f_{\mathrm PS}$. In spite of these formal remarks,
it is possible, however,  that the fact that the neutral pion is lighter
than the charged one (by about 20\% at $a\mu = 0.0040$, see 
section~(\ref{EIB})) makes inadequate the continuum $\chi$PT description
of finite size effects adopted in the present analysis. This caveat 
represents a further
motivation for simulations on larger lattices, which will eventually resolve
the issue.

\noindent (iv) Extrapolation to physical quark masses. We are assuming that 
$\chi$PT at next to leading order for the $N_f=2$ case is appropriate 
to describe the quark mass dependence of $m_{\mathrm PS}^2$ and $f_{\mathrm PS}$
up to $\sim 450$--$500$~MeV. Our lattice data are consistent with this, 
but it would be useful to include higher order terms in the $\chi$PT fits 
as well as more values of $a\mu$ to check this assumption. 
The effect of strange quarks in the sea should also be explored. 

\subsection{Effects of Isospin Breaking}
\label{EIB}

In this section we  report the results of some quantitative 
investigation of the effects of isospin breaking in 
the twisted mass formulation of lattice QCD at finite lattice spacing. This effect is
expected to be largest in the mass splitting between the lightest
charged and uncharged pseudo scalar mesons. A first analysis
at $a\mu=0.004$, taking the disconnected contribution in the neutral
channel fully into account, shows that the uncharged pseudo scalar
meson is about $20\%$ lighter than the charged one. We obtain
\[
  am_\mathrm{PS}^\pm = 0.1359(7)\, ,\qquad am_\mathrm{PS}^0 = 0.111(11)\, ,
\]
or, expressed differently, 
$r_0^2((m_\mathrm{PS}^0)^2-(m_\mathrm{PS}^\pm)^2)=c(a/r_0)^2$ 
with $c=-4.5(1.8)$. 
This coefficient is a factor of 2 smaller than the value found in
quenched investigations~\cite{Farchioni:2005hf}. 
Note that the uncharged pion being lighter than the charged one is
compatible with predictions from lattice $\chi$PT if the first order
phase transition scenario is realised \cite{Munster:2004am,Scorzato:2004da,Sharpe:2004ps}. 
For an investigation of isospin breaking effects in $\chi$PT see also
Ref.~\cite{Walker-Loud:2005bt}.

The disconnected correlations needed for the $\pi^0$ meson are evaluated
using a stochastic (Gaussian)  volume source with 4 levels of
hopping-parameter variance reduction \cite{McNeile:2000xx}. We use $24$
stochastic sources per gauge configuration and evaluate 
the relevant propagators every $10$-th
trajectory.

\section{Summary}

In this letter we 
have presented results of simulations 
of lattice QCD with $N_f=2$ maximally twisted Wilson quarks 
at a fixed value of the lattice spacing
$a\lesssim 0.1\ \mathrm{fm}$. 
We reached a pseudo scalar meson mass of 
about $300\ \mathrm{MeV}$. 
The numerical stability and smoothness of the simulations 
allowed us 
to obtain precise results for the pseudo scalar mass and 
decay constant which in turn led to determine 
the low energy constants of the effective chiral Lagrangian. 
In particular, we find for the pseudo scalar decay
constant in the chiral limit $F=121.3(7)\ \mathrm{MeV}$, 
and  $\bar{l}_3=3.65(12)$ and $\bar{l}_4=4.52(6)$
where only statistical errors are given.

We do see effects of isospin breaking which are largest in the mass splitting 
of the neutral and charged pions and turn out to be about 20\%. This is 
significantly  
smaller and opposite in sign 
than the corresponding splitting obtained in the quenched 
approximation. 

Tuning to maximal twist had to be performed on lattices of the same size 
as those used for the calculation of physical 
quantities. The reason for this is that we need to 
single out cleanly the one pion sector in order to impose the 
vanishing of the PCAC quark mass (Eq.~(\ref{eq:mpcac})) 
without being affected 
by finite size effects or excited state contributions.
Thus the tuning step itself is rather expensive. But it has to be 
done only once, as is the case for the determination of action
improvement coefficients in other Wilson based approaches.
Note, however, that with twisted mass fermions we do not 
have to compute further operator-specific improvement coefficients.

The encouraging results presented here will be extended and checked 
by future simulations that will cover one coarser and one finer 
lattice spacing, double the statistics at one of our present simulation
points ($\beta=3.9, a\mu=0.004$) 
and go to a larger, $32^3\cdot 64$, volume at the latter simulation point.
In this way, we will be able to obtain results in the 
continuum limit, cross-check 
our autocorrelation times, improve 
our error estimates and control 
the finite size effects in order to check $\chi$PT predictions. The
preliminary results presented in Refs.~\cite{Jansen:2006rf,Shindler:2006tm}
indicate a very good scaling behaviour already suggesting that 
automatic $\mathcal{O}(a)$ improvement is indeed working well. 

\subsubsection*{Acknowledgments}

The  computer time for this project was made available to us by the
John von Neumann-Institute for Computing on the JUMP and  Jubl systems
in J\"ulich and apeNEXT system in Zeuthen, by UKQCD
on the QCDOC machine at Edinburgh, by INFN on the apeNEXT systems in Rome, 
by BSC on MareNostrum in Barcelona (www.bsc.es)
and by the Leibniz Computer centre in Munich on the Altix system.  We
thank these computer centres and their staff for all technical  advice
and help. 
On QCDOC we have made use of Chroma~\cite{Edwards:2004sx} and BAGEL~\cite{BAGEL}
software and we thank members of UKQCD for assistance. For the analysis we used
among others the R language for statistical computing \cite{R:2005}.
We gratefully acknowledge discussions with D.~Be\'cirevi\'c, B.~Blossier,
N.~Christian, G.~M{\"u}nster, O.~P\`ene and A.~Vladikas.

This work has been supported in part by  the DFG 
Sonder\-for\-schungs\-be\-reich/Transregio SFB/TR9-03  and the EU Integrated
Infrastructure Initiative Hadron Physics (I3HP) under contract
RII3-CT-2004-506078.  We also thank the DEISA Consortium (co-funded by
the EU, FP6 project 508830), for support within the DEISA Extreme
Computing Initiative (www.deisa.org).  G.C.R. and R.F. thank MIUR (Italy)
for partial financial support under  the contract PRIN04. 
V.G. and D.P. thank MEC (Spain) for partial financial support under grant 
FPA2005-00711.

\bibliographystyle{h-physrev4}
\bibliography{paper}

\end{document}